\documentclass[showpacs,preprintnumbers,prd,nofootinbib,floats,amssymb,floatfix]{revtex4}
\usepackage{amsmath}
\usepackage{graphicx}
\usepackage{amsxtra}
\usepackage{hyperref}
\usepackage{amsmath}
\usepackage{amstext}
\begin{document}
\title{ The Hamilton-Jacobi analysis and Canonical Covariant description for three dimensional Palatini theory plus a Chern-Simons term  }

\author{Alberto Escalante}  \email{aescalan@ifuap.buap.mx}
\author{ Aldair-Pantoja}  \email{jpantoja@ifuap.buap.mx}
 \affiliation{  Instituto de F{\'i}sica, Benem\'erita Universidad Aut\'onoma de Puebla. \\
 Apartado Postal J-48 72570, Puebla Pue., M\'exico, }
\begin{abstract}
By using the Hamilton-Jacobi [HJ] framework the three dimensional Palatini theory plus a Chern-Simons term [PCS] is analyzed. We report the complete set of $HJ$  Hamiltonians  and a generalized $HJ$ differential from which all  symmetries of the theory are identified. Moreover,   we show that in spite of   PCS Lagrangian produces  Einstein's equations,   the generalized $HJ$ brackets  depend on a  Barbero-Immirzi like parameter. In addition we complete our study by performing a canonical covariant analysis, and we construct a closed and gauge invariant two form that encodes the symplectic geometry of the covariant phase space. 
\end{abstract}
 \date{\today}
\preprint{}
\maketitle
\section{Introduction}
Nowadays, the analysis of singular systems  has been the  cornerstone for studying  all fundamental forces  in nature. From the standard model, string theory,  to canonical gravity and Loop Quantum Gravity there is a big effort for understanding the underlying symmetries of  these systems \cite{ 2, 3, 4, 5, 6}.   In fact,  these forces expose  symmetries  and it is mandatory to perform   the study  of these symmetries  by using   alternative frameworks beyond standard classical mechanics. In this respect,  we can cite several approaches such as  the Dirac-Bergman, Faddeev-Jackiw, Canonical Covariant and the Hamilton-Jacobi methods \cite{7, 8, F2, F3, F4, F5, F6, F7, F8, F9, F10, F11, F12, F13, F14, F15, F16, F17, F18, F19, F20}. The Dirac approach allows us to identify the constraints of singular systems, which are classified into first class and second class. The formed  are generators of the gauge symmetry  and the latter   are used for constructing the Dirac brackets of the theory;  with the constraints at hand the symmetries of the theory can be identified.  Nonetheless,   the classification between the constraints into  first or  second class  is a difficult task, and alternative  approaches can be required. In this respect,  the Faddeev-Jackiw framework  allows  the construction of a symplectic tensor from which the symmetries of the theory can be identified. In the FJ framework it is not necessary to perform the classification of the constraints as in Dirac's method is done;  for gauge systems in order to obtain the symplectic tensor   it is  necessary fixing  the gauge,  and this fact could  complicate the   analysis. On the other hand, the canonical covariant method is a symplectic approach  based on the construction of  a closed  and gauge invariant  symplectic two form. From the symplectic two form we can perform a Hamiltonian analysis, however, in this approach we have not control on the constraints of the theory and relevant information of the symmetries can be missed;  in addition, the constraints are useful for performing the counting of  physical  degrees of freedom, hence this step can not be carryout.   Alternatively,  the HJ approach developed by  G\"uler  is based on the construction of a fundamental differential defined on the phase space, and the  fundamental blocks are   the constraints of the theory called Hamiltonians. The HJ Hamiltonians  can be involutives or noninvolutives and they are fundamental  blocks for obtaining   the characteristic equations, the gauge symmetries  and  the  generalized HJ brackets. The construction of the fundamental differential is direct and  the process for identifying the symmetries is in general more economical than in  the other approaches;  in this sense the HJ  framework is an interesting alternative for analyzing gauge systems. \\
Along  the ideas exposed above, the fundamental subject  of this paper is to report the HJ and canonical covariant analysis for  3d gravity described  in terms of Palatini's theory plus a Chern-Simons term [PCS] coupled through  an arbitrary Immirzi-like parameter  called $\gamma$ \cite{f21}.  It is well-known that the addition of topological terms to physical actions   does  not modify the equations of motion,  but there is a modification  on the fundamental brackets; in this respect two theories sharing the same classical equations of motion do not are equivalents at all \cite{f22, f23, f24}. This fact is present in the four dimensional Holst action,  described by Palatini's theory plus the addition of a topological term, the so-called  Holst term \cite{f25}. In this respect, the equations of motion of Palatini's theory and Holst theory are the same, however, at Hamiltonian level the structures  of the constraints of these theories are  different;  other examples   concerned to  this respect   can be consulted in the following references \cite{f26, f27, f28, f29}. Hence,  we have  a similar scenario for PCS theory;  three dimensional Palatini and PCS theories share the same equations of motion, the fundamental brackets, however, are different. We  use   the  G\"uler-HJ approach \cite{F17, F18, F19, F20, F21a}  because it is   an elegant and economical framework for analyzing  singular systems; in fact, we will extend  those results reported in \cite{ f29,  f30}. On the other hand, we want to report alternative studies  beyond  Dirac's and Faddeev-Jackiw framework in order to   have the best alternative  for analyzing singular systems. \\
The   paper is organized as follows. In the Section II we develop the HJ analysis for PCS theory. We construct a fundamental differential where the characteristics equations and all symmetries of the theory are found. We reproduce and extend the results reported in \cite{f29, f30}. In Section III the canonical covariant formalism is performed; we construct a closed and gauge invariant geometric structure from which  a Hamiltonian description of the theory is developed, and we identify the symmetries of the theory, however, we comment the disadvantages of this formalism with  respect the other ones reported in the literature.

\section{Hamilton-Jacobi analysis}
We start with the following action expressed  as   Palatini's  3d gravity theory plus a Chern Simons term \cite{f21}
\begin{equation}
S[e,A]=2\int_{M} e^{i}\wedge F_{i} + \frac{1}{\gamma}\int_{M}[2A^{i}\wedge dA_{i}+\frac{2}{3}\varepsilon_{ijk}A^{i}\wedge A^{j}\wedge A^{k}], 
\label{ac1}
\end{equation}
where $M$ is a three dimensional manifold without boundary, $F_i= dA_i + \frac{1}{2}\epsilon_{ijk}A^j\wedge A^k $ is the strength curvature  of the 1- form connexion $A^i$, the $e$'s  are  the triad fields,  $i,j,k .. = 0,1,2$ are internal $SU(2)$ indices,  and     $\gamma$ is a  Barbero-Immirzi-like  parameter.  From the variation of the action,  the following equations of motion arise 
\begin{eqnarray}
\epsilon^{\alpha \mu \nu }(D_\mu e^i_\nu + \frac{1}{\gamma}F^i_{\mu \nu}) &=& 0, \nonumber \\
\epsilon^{\alpha \mu \nu } F^i_{\mu \nu} &=&0,
\label{eqm}
\end{eqnarray}
where $D_\mu e^i_\nu= \partial_\mu e^i_\nu+ \varepsilon^{i}{_{jk}} A^j_\mu e^k_\nu $. These equations for different values of $\gamma$ represent a set of  equations   classically equivalent to   three dimensional  Einstein's  theory, however,  in spite of this equivalence  we will see that the generalized $HJ$ brackets depend on the $\gamma$ parameter, while in Palatini theory there is not such  a dependence \cite{f31}, in this sense the Palatini theory and PCS are different to each other. Along the paper, we will use the notation $\mu, \nu, ... = 0,1,2$ for spacetime indices and the alphabet letters  $a, b, c$ for space indices. Moreover,  we will suppose that the manifold has topology $M= \Sigma \times R $, where $\Sigma$ is a Cauchy surface and $R$ is an evolution parameter.  With these considerations at hand we  perform    the 2+1 decomposition,   and the action $(\ref{ac1})$ takes the following form
\begin{eqnarray}
S[e,A] &=& \int \epsilon^{ab}\left[ \left(e_{b}^{\;\;i} + \frac{1}{\gamma}A_{b}^{\;\;i} \right)\dot{A}_{ai} + \frac{1}{2}e_{0}^{\;\;i}F_{abi} + A_{0i}\left( D_{a}e_{b}^{\;\;i} + \frac{1}{\gamma}F_{ab}^{\;\;i}   \right)     \right] d^{3}x,  
\label{S0}
\end{eqnarray}
we have removed an overall factor of $2$ which does  not affect to the equations of motion and we have defined $\epsilon^{0ab}\equiv \epsilon^{ab}$. Moreover,  here we have used 
\begin{eqnarray*}
F_{abi}&=&\partial_{a}A_{bi} - \partial_{b}A_{ai} + \varepsilon _{i}^{\;\;jk}A_{aj}A_{bk},\\
D_{a}e_{bi} &=& \partial_{a}e_{bi} + \varepsilon_{i}^{\;\;jk}A_{aj}e_{bk}.
\end{eqnarray*}
The action (\ref{S0}) has been analyzed by using the Dirac and Faddeev-Jackiw approaches in \cite{f21, f30}; in these works the Dirac and Faddeev-Jackiw  constraints,  a symplectic tensor and the symmetries of the theory  were reported. However, in the present  paper we will extend those works by performing a $HJ$ analysis and we will reproduce in more economical way those results.   Furthermore, we identify the canonical momenta $(\Pi_ i^{\mu}, p^{\mu}{_{i}})$ conjugated to $(A^i_{\mu}, e^{i}{_{\mu}})$ 
\begin{align*}
\Pi_ i^{\mu} &\equiv \frac{\partial\mathcal{L}}{\partial\dot{A}^i_{\mu}}, & 	
p^{\mu}{_{i}} \equiv  \frac{\partial\mathcal{L}}{\partial\dot{e}^{i}{_{\mu}}},  
\end{align*}
Hence, by using the momenta, from  the action (\ref{S0}) we identify the following $HJ$ Hamiltonians
\begin{eqnarray}
H' &\equiv& \Pi+H_{0} =0, \nonumber \\
\phi_{i} &\equiv& p_{i}^{0}=0, \nonumber \\
\tilde{\phi}_{i} &\equiv&\Pi_{i}^{0}=0, \nonumber \\
\varphi_{i}^{a} &\equiv& p_{i}^{a} =0, \nonumber \\
\tilde{\varphi}_{i}^{a}&\equiv&  \Pi_{i}^{a} -\epsilon^{ab}(e_{bi}+\frac{1}{\gamma}A_{bi})=0,
\label{hamil1}
\end{eqnarray}
where $\Pi= \partial_0 S$,  identifying to $S$ with the action and $H_0$  with the canonical Hamiltonian 
\begin{equation}
H_0= -\frac{\epsilon^{ab}}{2}e^i_0F_{abi} - A_0^i (D_a \Pi^a_i+ \frac{1}{\gamma} \epsilon^{ab} \partial_a A_{bi}). 
\end{equation}
the definition of the momenta allows  us to identify  the fundamental Poisson brackets 
\begin{eqnarray}
\lbrace e_{\alpha}^{i}(x),p_{j}^{\mu}(y) \rbrace &=& \delta_{\alpha}^{\mu}\delta_{j}^{i}\delta^{2}(x-y), \nonumber \\
\lbrace A_{\alpha}^{i}(x), \Pi_{j}^{\mu}(y) \rbrace &=& \delta_{\alpha}^{\mu}\delta_{j}^{i}\delta^{2}(x-y).
\end{eqnarray}
In this manner, with the  Hamiltonians identified, we construct the fundamental differential which describes the evolution of any  function, say $f$, on the phase space \cite{ F17, F18, F19, F20}
\begin{equation*}
df(x)=\int d^{2}y\left(
\lbrace f(x),H'(y) \rbrace dt + \lbrace f(x),\varphi_{i}^{a}(y) \rbrace d\xi_{a}^{i} + \lbrace f(x),\tilde{\varphi}_{i}^{a}(y) \rbrace d\tilde{\xi}_{a}^{i}+
\lbrace f(x),\phi_{i}(y) \rbrace d\lambda^{i} + \lbrace f(x),\tilde{\phi}_{i} \rbrace d\tilde{\lambda}^{i} \right), 
\end{equation*}
here, $\xi_{a}^{i}, \tilde{\xi}_{a}^{i}, \lambda^{i}$, and $ \tilde{\lambda}^{i}$ are parameters associated with  the Hamiltonians.  On the other hand, we observe that the Hamiltonians $\phi_{i}$ and $\tilde{\phi}_{i}$ are involutives and $\varphi_{i}^{a}$, $\tilde{\varphi}_{i}^{a}$ are non-involutives. Involutive Hamiltonians, are those whose Poisson brackets with all Hamiltonians,  including themself,    vanish; otherwise, they are called non-involutives. The presence of non-involutive Hamiltonians introduce the generalized $HJ$ brackets defined by \cite{ F17, F18, F19, F20}
\begin{align}
\{A, B\}^{*} &=\{A, B\} - \{A, H'_{\bar{a}}\}(C{_{\bar{a}\bar{b}}})^{-1}\{H'_{\bar{b}}, B\},
\label{10}
\end{align}
where $(C{_{\bar{a}\bar{b}}})$ is the matrix whose entries are given by  the Poisson brackets between non-involutives Hamiltonians and $(C{_{\bar{a}\bar{b}}})^{-1}$ its inverse;   that matrix takes the form
\begin{equation*}
C_{\alpha\beta}
=
\begin{pmatrix}
0 & \epsilon^{ab}\delta_{ij} \\
\\
\epsilon^{ba}\delta_{ij} & -\frac{2}{\gamma}\epsilon^{ab}\delta_{ij} \\
\end{pmatrix}
\delta^{2}(x-y),
\end{equation*}
with 
\

\begin{equation*}
C_{\alpha\beta}^{-1}
=
\begin{pmatrix}
-\frac{2}{\gamma}\epsilon_{ab}\delta^{ij} & \epsilon_{ab}\delta^{ij} \\
\\
-\epsilon_{ba}\delta^{ij} & 0 \\
\end{pmatrix}
\delta^{2}(x-y), 
\end{equation*}
hence, the generalized brackets between the fields read  
\begin{eqnarray}
\lbrace e_{a}^{i}(x),e_{b}^{j}(y) \rbrace^* &=& -\frac{2}{\gamma}\epsilon_{ab}\delta^{ij}\delta^{2}(x-y), \nonumber \\
\lbrace e_{a}^{i}(x),A_{b}^{j}(y) \rbrace^*  &=& \epsilon_{ab}\delta^{ij}\delta^{2}(x-y), \nonumber \\
\lbrace A_{a}^{i}(x),A_{b}^{j}(y) \rbrace^*  &=& 0, \nonumber \\
\lbrace e_{a}^{i}(x),\Pi_{j}^{b}(y) \rbrace^*  &=& -\frac{1}{\gamma}\delta_{a}^{b}\delta_{j}^{i}\delta^{2}(x-y), \nonumber \\
\lbrace A_{a}^{i}(x),\Pi_{j}^{b}(y) \rbrace^*  &=& \delta_{a}^{b}\delta_{j}^{i}\delta^{2}(x-y), \nonumber \\
\lbrace e_{a}^{i}(x),p_{j}^{b}(y) \rbrace^*  &=& 0, \nonumber \\
\lbrace \Pi_{a}^{i}(x),\Pi_{j}^{b}(y) \rbrace^* &=& 0, \nonumber 
\end{eqnarray}
note that the fields  $e'$s are noncommutative due to the presence  of the    $\gamma$  parameter.   This fact makes PCS  theory different to standard Palatini action where the triad is  commutative \cite{f31}.  With the generalized brackets at hand, we introduce the new  fundamental  $HJ$ differential 
\begin{equation}
df(x)=\int d^{2 }y \left[ \lbrace f(x),H'(y) \rbrace^* dt + \lbrace f(x),\phi_{i}(y) \rbrace^* d\lambda^{i} + \lbrace f(x), \tilde{\phi}_{i}(y) \rbrace^* d\tilde{\lambda}^{i} \right], 
\label{dif2}
\end{equation}
and  we observe that the non-involutive  Hamiltonians has been removed. On the other hand, the Frobenius integrability conditions on the Hamiltonians $\phi_{i}$ and  $\tilde{\phi}_{i}$ introduce new $HJ$ Hamiltonians. In fact, integrability conditions are relevant because ensure that the system (\ref{dif2}) is integrable. From the integrability conditions the following Hamiltonians arise 
\begin{eqnarray}
d\phi_{i}(x) =\int d^{2}y \Big[ \lbrace \phi_{i}(x)),H'(y) \rbrace dt + \lbrace \phi_{i}(x),\phi_{j}(y) \rbrace d\lambda^{j} + \lbrace \phi_{i}(x), \tilde{\phi}_{j}(y) \rbrace d\tilde{\lambda}^{j} \Big]=0, \nonumber \\
\rightarrow \tau_{i} \equiv \epsilon^{ab}F_{abi}=0, \nonumber \\
d\tilde{\phi}_{i}(x)=\int d^{2}y \left[ \lbrace \tilde{\phi}_{i}(x),H'(y) \rbrace dt + \lbrace \tilde{\phi}_{i}(x),\phi_{j}(y) \rbrace d\lambda^{j} + \lbrace \tilde{\phi}_{i}(x), \tilde{\phi}_{j}(y) \rbrace d\tilde{\lambda}^{j} \right]=0, \nonumber \\
\rightarrow \tilde{\tau}_{i} \equiv D_{a}\Pi_{i}^{a} + \frac{1}{\gamma} \epsilon^{ab}\partial_{a}A_{bi}=0.
\end{eqnarray}
The generalized algebra between the new Hamiltonians $\tau_i$ and $\tilde{\tau}_i$ is given by
\begin{eqnarray}
\lbrace \tau_{i}, \tau_{j} \rbrace ^{*}&=&0, \nonumber \\
\lbrace \tau_{i},\tilde{\tau}_{j} \rbrace ^{*}&=&\varepsilon_{ijk}{\tau}^{k} , \nonumber \\
\lbrace \tilde{\tau}_{i},\tilde{\tau}_{j} \rbrace ^{*} &=& \varepsilon_{ijk}\tilde{\tau}^{k}, 
\end{eqnarray}
where we can observe that  these  Hamiltonians are involutive, therefore we do not expect new  Hamiltnonians. Furthermore,  the Hamiltonians $\tau_{i}$ and $\tilde{\tau}_i$ form a Poincar\'e algebra. In fact, $\tau_i$ is related to translations and  $\tilde{\tau}_i$ is related to rotations. 
With all involutive Hamiltonians at hand we construct the following generalized differential
\begin{eqnarray}
df(x)&=&\int d^{2}y \Big[ \lbrace f(x), H'(y) \rbrace^* dt + \lbrace f(x), \phi_{i} \rbrace^* d\lambda^{i}(y) + \lbrace f(x), \tilde{\phi}_{i}(y) \rbrace^* d\tilde{\lambda}^{i} + \lbrace f(x),\tau_{i}(y) \rbrace^* d\Upsilon^{i}(y) \nonumber \\
&+& \lbrace f(x),\tilde{\tau}_{i}(y) \rbrace^* d\tilde{\Upsilon}^{j} \Big], 
\end{eqnarray}
where $\Upsilon^{i}$ and $\tilde{\Upsilon}^i$ are parameters related with the Hamiltonians  $\tau_{i}$ and $\tilde{\tau}_i$  respectively. In this manner, from the fundamental differential we can calculate  the characteristic equations \cite{ F17, F18, F19, F20},which will  reveal the symmetries of the theory. The characteristic equations are given by 
\begin{eqnarray}
de_{0}^{i}&=&d\lambda^{i},  \nonumber \\
dA_{0}^{i}&=&d\tilde{\lambda}^{i},  \nonumber \\
de_{a}^{i}&=&\left(\partial_{a}e_{0}^{i}+\varepsilon_{l}^{\;i}\;_{k}e_{0}^{l}A_{a}^{k}+\varepsilon_{l}^{\;i}\;_{k}A_{0}^{l}e_{a}^{k}\right)dt
-2D_{a}d{\Upsilon}^{i}-\varepsilon_{j}{^{il}}e_{al}d\tilde{\Upsilon}^{j}, \nonumber \\
dA_{a}^{i}&=&F_{a0}^{i}dt-D_{a}d\tilde{\Upsilon}^{i}.
\end{eqnarray}
hence, from the temporal part  we identify the equations of motion 
\begin{eqnarray}
\partial_{0}e_{a}^{i}&=&\partial_{a}e_{0}^{i}+\varepsilon_{l}^{\;i}\;_{k}e_{0}^{l}A_{a}^{k}+\varepsilon_{l}^{\;i}\;_{k}A_{0}^{l}e_{a}^{k}, \nonumber \\
\partial_{0}A_{a}^{i}&=&\partial_{a}A_{0}^{i}-\varepsilon_{lj}{^{i}}A_{0}^{l}A_{a}^{j}, 
\end{eqnarray}
which correspond  to  3d Einstein's equations. On the other hand, we observe from the characteristic equations that the fields $e_0^i$ and $A_0^i$  are  not related neither  $t$ nor $\Upsilon^i$ and $ \tilde{\Upsilon}^i$ parameters, which  means that they are   identified as Lagrange multipliers. Moreover, the parameters associated  with the involutives Hamiltonians are  related to  the following gauge transformations 
\begin{eqnarray}
\delta e_{a}^{i}=D_{a}\delta {\Upsilon}^{i} + \frac{1}{2}\varepsilon_{j}{^{il}}e_{al}\delta \tilde{\Upsilon}^{j}, \nonumber \\
\delta A_{a}^{i} = \frac{1}{2}D_{a}\delta \tilde{\Upsilon}^{i}.
\label{gauge}
\end{eqnarray}
We can observe that similar  results were reported in \cite{f30}, where different approaches were used, however, we can note  that the $HJ$ approach is an  economical way for finding the symmetries of the theory. \\
We finish this section by performing  the counting of physical degrees of freedom: in this formalism the physical degrees of freedom are identified   with the dynamical fields found    in the characteristic equations minus  the complete set of involutive Hamiltonians. For this theory, the dynamical  variables are the following six $e^i_a$ and six $A_a^i$;  the involutive   Hamiltonians are 12 $(\phi_{i}, \tilde{\phi}_{i}, \tau_{i}, \tilde{\tau}_{i} )$, and thus  $DF= 12-12=0$, the theory is devoid of physical degrees of freedom,  as expected. \\
In the following section, we will complete our analysis by performing the canonical covariant method of the PCS theory. 
\section{Canonical  covariant  analysis}
We start our study by taking the variation of the action (\ref{ac1}) with  respect the dynamical  fields, 
\begin{eqnarray}
\delta S[e, A]= \int_M \Big[ \Big( \varepsilon^{\alpha \mu \nu} (D_{\mu} e_{\nu}^{i}+\frac{1}{\gamma}F_{\mu\nu}^{i} ) \Big) \delta A_{\alpha i} + \Big(  \varepsilon^{\alpha\mu\nu}F_{\mu\nu i} \Big) \delta e^i_\alpha + \partial_\mu \Big( \varepsilon^{\mu\alpha\nu} (e_{\alpha}^{i}+\frac{1}{\gamma}A_{\alpha}^{i} )\delta A_{\nu i} \Big)\Big] dx^3, 
\end{eqnarray}
where we identify the equations of motion (\ref{eqm}) and from the divergence term a symplectic potential is identified   \cite{F16}
\begin{equation}
\Psi^{\mu}=\varepsilon^{\mu\alpha\nu}\left(e_{\alpha}^{i}+\frac{1}{\gamma}A_{\alpha}^{i}\right)\delta A_{\nu i}.
\end{equation}
In this manner,  we define  the essential  object   of the canonical covariant method, the covariant phase space; the covariant phase space for the theory described by (\ref{ac1}) is the space of solutions of the equations of motion (\ref{eqm}), and we will call it $Z$ \cite{F16, f23}. Hence, on $Z$ the fields $A_\mu ^i$ and $e_\mu^i$ are zero-forms and its variations (exterior derivation on $Z$) $\delta A_\mu ^i $ and $\delta e_\mu^i$ are 1-forms. Therefore, the variation of the symplectic potential generates the two form symplectic structure,   
\begin{equation}
\omega=\int_{\Sigma}J^{\mu}d\Sigma_{\mu}=\int_{\Sigma}\delta\Psi^{\mu}d\Sigma_{\mu}=\int_{\Sigma}\varepsilon^{\mu\alpha\nu}\delta \left(e_{\alpha}^{i}+\frac{1}{\gamma}A_{\alpha}^{i}\right)\wedge\delta A_{\nu i} d\Sigma_{\mu}. 
\end{equation}
where $\Sigma$ is a Cauchy surface.  We will find the symmetries of the theory trough that  geometric structure. In fact, we will prove that $\omega$ is closed and gauge invariant; the closeness of $\omega$ is equivalent  to the Jacobi identity that Poisson brackets satisfy in the  Hamiltonian scheme. In addition, we know that  gauge invariance is reflection  of an internal  symmetry  when  the theory is singular.     Furthermore, the integral kernel $J^{\mu}$ of the geometric form, is conserved; this fact will be  important because it guarantees that $\omega$ is independient of $\Sigma$. Hence, we observe that $\delta^{2}e^{i}{_{\mu}} = 0$ and $\delta^{2}A_{\alpha}^i=0$, due to  $e^{i}{_{\mu}}$ and $A_{\alpha}^i$ are independent zero forms on $Z$ and $\delta$ is nilpotent, therefore $\omega$ is closed. Now, we shall find  the linearized equations of motion;   they are obtained from the substitutions    $A_\mu ^i$ $\rightarrow$ $A_\mu ^i + \delta A_\mu^i$, and $e_\mu^i  \rightarrow e_\mu^i + \delta e_\mu^i$ into the equations of motion,  and keeping only the first order terms, hence 
\begin{eqnarray}
\epsilon^{\alpha\mu\nu}D_{\mu}\delta A_{\nu i}=0, \nonumber \\
\epsilon^{\alpha\mu\nu}\left(D_{\mu}\delta A_{\nu}^{i}+\varepsilon_{ijk}\delta A_{\mu}^{j}e_{\nu}^{k}\right)=0, 
\label{lin}
\end{eqnarray}
the linearized  equations  will be important for proving the conservation of $J^\mu$. In fact,  by taking the generator of rotations in  the gauge transformations (\ref{gauge}) and  under an arbitrary variation we obtain 
\begin{eqnarray}
\delta e_{\mu}^{'i}=\delta e_{\mu}^{i} + \frac{1}{2}\varepsilon_{j}{^{il}}\delta e_{\mu l}\epsilon^{j}, \nonumber  \\
\delta A_{\mu}^{'i}=\delta A_{\mu}^{i} + \frac{1}{2}\varepsilon^{i}{_{jk}}\delta A_{\mu}^{j}\epsilon^{k},
\label{gau2}
\end{eqnarray}
where we have called $\epsilon^i \equiv d\Upsilon^{i}$ and $\tilde{\epsilon}^i \equiv d\tilde{\Upsilon}^{i}$.  In this manner, under the transformations (\ref{gau2}) the symplectic structure transforms as 
\begin{eqnarray}
\omega'&=&\int_\Sigma \left(\varepsilon^{\mu\alpha\nu}\delta e_{\alpha}^{'i}\wedge\delta A_{\nu i}^{'} + \frac{1}{\gamma}\varepsilon^{\mu\alpha\nu}\delta A_{\alpha}^{' i}\wedge\delta A_{\nu i}^{'}\right)d\Sigma_{\mu}, \nonumber \\
&=&\omega+\int_{\Sigma}O(\epsilon^{2}), 
\end{eqnarray}
thus, $\omega$ is a $SU(2)$ singlet. Hence, this fact allows us prove  the conservation of $J^\mu $, this is 
\begin{eqnarray}
\partial_{\mu}J^{\mu}&=&D_{\mu}J^{\mu}, \nonumber \\
&=&\epsilon^{\mu\alpha\nu}D_{\mu}\delta e_{\alpha}^{i}\wedge\delta A_{\nu i} + \epsilon^{\mu\alpha\nu}\delta e_{\alpha}^{i}\wedge D_{\mu}\delta A_{\nu i} +\frac{1}{\gamma} \epsilon^{\mu\alpha\nu}D_{\mu}\delta A_{\alpha}^{i}\wedge\delta A_{\nu i} + \frac{1}{\gamma}\epsilon^{\mu\alpha\nu}\delta A_{\alpha}^{i}\wedge D_{\mu} \delta a_{\nu i},  \nonumber \\
&=&0
\end{eqnarray}
where we have used the linearized equations of motion (\ref{lin}) and the antisymmetry  of the 1-forms $\delta e_\mu^i$ and $\delta A_a^i$.\\
On the other hand,  we know that both Palatini and Chern-Simons theories  are  diffeomorphism covariant, and this important symmetry must to be contained in the fundamental gauge transformations. In fact,  with the particular choice  
\begin{equation*}
\delta \Upsilon^{i}=\epsilon^{\rho}e_{\rho}^{i} \quad \mathrm{and} \quad \delta \tilde{\Upsilon}^{i}=2\epsilon^{\mu}A_{\mu}^{i}, 
\end{equation*}
from the gauge transformations (\ref{gauge})  we obtain 
\begin{eqnarray}
\delta e_{\alpha}^{' i} &=& \delta e_{\alpha}^{i} + \mathcal{L}_{\vec{\epsilon}}\delta e_{\alpha}^{i}, \nonumber \\
\delta A_{\alpha}^{'i}&=&\delta A_{\alpha}^{i} + \mathcal{L}_{\vec{\epsilon}}\delta A_{\alpha}^{i}. 
\label{diff}
\end{eqnarray}
this means that diffeomorphisms  are  identified as  internal symmetry of the theory. In this manner, we can prove that the symplectic structure transforms under (\ref{diff}) as
\begin{eqnarray}
\omega^{'}&=&\int_{\Sigma}\varepsilon^{\alpha\beta\mu}\left(\delta e_{\beta}^{' i}+\frac{1}{\gamma}\delta A_{\beta}^{' i}\right)\wedge \delta A_{\mu i}^{'}d\Sigma_{\alpha}, \nonumber \\
&=&\omega + \int_{\Sigma}\mathcal{L}_{\vec{{\epsilon}}}\omega
\end{eqnarray} 
however, $\mathcal{L}_{\vec{\epsilon}}\omega = \vec{\epsilon}\cdot d\omega + d(\vec{\epsilon}\cdot\omega)$, but $\delta\omega=0$ (it is closed) and, the term $d(\vec{\epsilon}\cdot\omega)$ corresponds  to a surface term. Therefore $\omega$ is invariant under infinitesimal diffeomorphisms. \\
Once we have found the symmetries of the theory from the symplectic point of view, we  consider that upon picking $\Sigma$ to be the standard initial value surface $t=0$, the symplectic structure takes the standard  form  
\begin{equation}
\omega=\int_{\Sigma} \delta\Pi_{i}^{a}\wedge\delta A_{a}^{i}, 
\label{simp}
\end{equation}
where  $\Pi_{i}^{a}=\varepsilon^{ba}\left(e_{ib}+\frac{1}{\gamma}A_{bi}\right)$.  In this manner, under these considerations, we are able to perform a Hamiltonian study. In fact,  let us consider to $f$ as  any 0-form  defined on $Z$, hence the Hamiltonian vector field defined by the symplectic form (\ref{simp}) is given by
\begin{align}
X_{f}\equiv\int_{\Sigma}\frac{\delta f}{\delta\Pi^{a}_i}\frac{\delta}{A_{a}^i} - \frac{\delta f}{\delta A^i_{a}}\frac{\delta}{\delta\Pi^{a}_i}. 
\label{vec}
\end{align}
Moreover, the Poisson bracket between two zero-forms is defined as usual
\begin{equation}
\{f,g\}_P\equiv -X_f g = \int_{\Sigma}\frac{\delta f}{A^i_{a}}\frac{\delta g}{\delta \Pi^{a}_i} - \frac{\delta f}{\delta \Pi^{a}_i} \frac{\delta g}{\delta A^i_{a}}.
\end{equation}
Then, if we  smearing the constraints with test fields, namely 
\begin{eqnarray}
\tau[N^{i}]=\int_{\Sigma}N^{i}\left[\varepsilon^{ab}F_{abi}\right], \\ \nonumber 
\tilde{\tau}[M^{i}]=\int_{\Sigma}M^{i}\left[D_{a}\Pi_{i}^{a}+\frac{1}{\gamma}\varepsilon^{ab}\partial_{a}A_{bi}\right], 
\end{eqnarray}
and we calculate the fundamental variations of these constraints 
\begin{equation*}
\frac{\delta \tau}{\delta A_{a}^{i}}=2\varepsilon^{ab}\partial_{b}N_{i}+2\varepsilon^{k}_{\;il}\varepsilon^{ab}A_{b}^{l}N_{k}, \quad \frac{\delta \tau}{\delta\Pi_{i}^{a}}=0, 
\end{equation*}
and 
\begin{equation*}
\frac{\delta \tilde{\tau}}{\delta A_{a}^{i}}=M^{k}\varepsilon_{ik}^{\;j}\Pi_{i}^{a}+\frac{1}{\gamma}\varepsilon^{ab}\partial_{b}M^{i}, \quad \frac{\delta \tilde{\tau}}{\delta \Pi_{i}^{a}}=-D_{a}M^{i}, 
\end{equation*}
this allows  us to calculate  the following Poisson brackets between the constraints and the fields 
\begin{eqnarray}
\lbrace A_{a}^{i}, \tau \rbrace &=&0, \nonumber \\
\lbrace A_{a}^{i},\tilde{\tau} \rbrace &=&-D_{a}M^{i},  \nonumber \\
\lbrace \Pi_{i}^{a}, \tau \rbrace &=& -\varepsilon^{ab}\partial_{b}N_{i}-\varepsilon^{k}_{\;il}\varepsilon^{ab}A_{b}^{l}N_{k}, \nonumber \\
\lbrace \Pi_{i}^{a},\tilde{\tau} \rbrace &=& -M^{k}\varepsilon_{ki}{^{j}}\Pi_{j}^{a}-\frac{1}{\gamma}\varepsilon^{ab}\partial_{b}M_{i}. 
\label{covp}
\end{eqnarray}
In this manner, by taking into account the equations (\ref{covp}), we observe that the motion generated by $\tau[N]$ and $\tilde{\tau}_{i}[M]$ is
\begin{eqnarray}
A_{a}^{' i} &\longrightarrow& A_{a}^{i} + D_{a}M^{i}\epsilon, \nonumber \\ 
\Pi_{i}^{' a} &\longrightarrow& \Pi_{i}^{a} + M^{k}\varepsilon_{ki}{^{\;j}}\Pi_{j}^{a}\epsilon +\frac{\epsilon}{\gamma} \varepsilon^{ab}\partial_{b}M^{i}+ \epsilon \epsilon^{ab}D_bN_i, 
\end{eqnarray}
where $\epsilon$ corresponds to  an infinitesimal parameter \cite{ F16, f23}. The  transformations of the connection  are those found by means the $HJ$ approach. However, it is important to comment some differences between the approaches used in this paper. We can note that in the $HJ$ method the dynamical variables  are given by the connection and the triad  fields,  in this sense,  the $HJ$ method is similar to perform a pure Dirac's  method \cite{f31} in which the   canonical momenta are associated with   all dynamical variables.  On the other hand,   in the canonical covariant method  the dynamical variables are those occurring   in the action with time derivative (see the action (\ref{S0})) and only  those variables are  associated with its canonical momenta. In this respect, in the canonical covariant method we will not find any  gauge transformations associated with  the triad field,  since  from the beginning it  is not a dynamical variable.
In addition, in order to calculate the gauge transformations  we used  the constraints found in the $HJ$ approach. In fact, we know that in the canonical covariant method we have not control on the constraints and this fact restricts  us to perform  the counting of physical degrees of freedom and the construction of  a generalized bracket such as in the $HJ$ scheme  is done. 
\section{Conclussions}
A detailed $HJ$ and canonical covariant analysis for $PCS$ theory were  developed. With  respect to the $HJ$ study,  we have constructed a generalized differential given in terms of the generalized brackets and involutive Hamiltonians allowing us identify the characteristic equations of the theory. The contribution of the   $\gamma$ parameter  is observed in the generalized brackets:  the triad becomes to be noncommutative and this fact makes   PCS  different at classical level  from   Palatini theory. Moreover,  the gauge transformations were reported and the counting of physical degrees of freedom was performed. 
On the other hand, from the symplectic point of view, a closed and gauge invariant geometric structure was constructed  and   the symmetries of the theory were identified.  We would point out  that  in this formalism we have  not control on the constraints of the theory, and this fact do not  allow us  to construct a kind of generalized brackets such as in the $HJ$  scheme  is done; in this respect the contribution of $\gamma$ is missed. In this manner, the $HJ$ framework show advantages with respect to the canonical covariant formalism. Therefore,  we have extended the results  presented in \cite{f21, f30} where  different approaches were used. \\
\noindent \textbf{Acknowledgements}\\[1ex]
 We would like to  thank R. Cartas-Fuentevilla for discussion on the subject and reading of the manuscript.

\end{document}